\documentclass[prd, twocolumn, nofootinbib, floatfix]{revtex4}

\usepackage{amsmath}
\usepackage{amsfonts}
\usepackage{graphicx}
\usepackage{hyperref}
\usepackage{color}
\input epsf

\newcommand{\beq}{\begin{equation}}
\newcommand{\eeq}{\end{equation}}
\newcommand{\beqa}{\begin{eqnarray}}
\newcommand{\eeqa}{\end{eqnarray}}

\def\imag{\dot \imath}
\def\pa{{\partial}}

\begin{document}
\title{Induced Gravity and the Attractor Dynamics of Dark Energy/Dark Matter}

\author{Jorge L.~Cervantes-Cota}
\email{jorge.cervantes@inin.gob.mx}
\affiliation{Depto.~de F\'{\i}sica, Instituto Nacional de
Investigaciones Nucleares, M\'{e}xico}
\affiliation{Berkeley Center for Cosmological Physics, University of California, Berkeley, CA, USA}
\author{Roland de Putter}
\email{rdeputter@berkeley.edu}
\affiliation{Berkeley Center for Cosmological Physics, University of California, Berkeley, CA, USA}
\affiliation{IFIC, Universidad de Valencia-CSIC, Valencia, Spain}
\affiliation{Institut de Ciencies del Cosmos, Barcelona, Spain}
\author{Eric V.~Linder}
\email{evlinder@lbl.gov}
\affiliation{Berkeley Center for Cosmological Physics, University of California, Berkeley, CA, USA} 
\affiliation{Institute for the Early Universe, Ewha Womans University, 
Seoul, Korea}

\begin{abstract}
Attractor solutions that give dynamical reasons for dark energy to 
act like the cosmological constant, or behavior close to it, are 
interesting possibilities to explain cosmic acceleration.  Coupling 
the scalar field to matter or to gravity enlarges the dynamical behavior; 
we consider both couplings together, which can ameliorate some problems 
for each individually.  Such theories have also been proposed in a 
Higgs-like fashion to induce gravity and unify dark energy and dark 
matter origins.  We explore restrictions on such theories due to their 
dynamical behavior compared to observations of the cosmic expansion. 
Quartic potentials in particular have viable stability properties and 
asymptotically approach general relativity. 
\end{abstract}

\date{\today}
\pacs{98.80.Cq, 98.80.-k, 04.50.Kd}
\preprint{}
\maketitle


\section{Introduction}
Over a decade ago two supernovae groups, the Supernova Cosmology Project 
and the High-Z Supernovae Search Team, provided evidence for an 
accelerated expansion of the Universe \cite{supernovae}. 
In recent years this discovery has gained more
evidence from a variety of observations: further supernova data 
\cite{union2}, measurements of 
the cosmic microwave background radiation \cite{wmap7} and  galaxy surveys
\cite{gal_surveys}.   One possibility to explain this acceleration is to 
introduce a new component within the dynamics of General Relativity (GR), 
either as a uniform constant (cosmological constant) or as a scalar field 
evolving along 
a potential (as in the inflationary scenario).  Since the current 
acceleration seems to be a unique phenomenon \cite{uniqcmb}, at least 
since the time of primordial nucleosynthesis, this poses a fine tuning 
problem unless some dynamical attractor exists. 

A different possibility is to look for acceleration as arising from 
modifications of gravity.  In many cases this can be viewed as coupling 
scalar fields nonminimally to gravity, within the framework of scalar-tensor 
theories \cite{BrDi61}, an approach called extended quintessence 
\cite{PeBaMa00}.   Attractor mechanisms can work here as 
well \cite{BaPi00,BaMaPe00,Ch01,FaJe06}, and also  in the case where a scalar 
field is coupled to (dark) matter \cite{Am00,DaCoKh06,BeFlLaTr08,Co08,QuBrBaPi08}.

In the present work we investigate the influence of both couplings 
on the attractor dynamics.  From a phenomenological point of view this 
enriches the phase space and also can help with problems that arise from 
one coupling or the other.  By comparison with observations of the 
cosmic expansion behavior we can constrain the allowed parameter space.  
From a theoretical point of view several models can lend motivation to 
such a combination of couplings. 

Induced gravity \cite{ig_old} is similar to standard scalar-tensor 
theories, but gravity is induced by a Higgs-like field.  
One motivation stems from Einstein's original ideas 
to incorporate Mach's principle into GR, by which the mass of a particle 
should originate from the interaction with all the particles of the 
universe, and so the interaction should be the gravitational one since 
it couples to all particles, i.e.\ to their masses or energies.  
To realize a stronger relationship with the material contents, Brans and 
Dicke \cite{BrDi61} introduced their scalar-tensor theory of gravity, making 
the gravitational coupling, that is Newton's constant, a scalar function 
determined by the distribution of the cosmic content. 
 
On the other hand, in modern particle physics the inertial mass is 
generated by the interaction with the Higgs field; the successful 
Higgs mechanism also lies precisely in the direction of Einstein's idea of
producing mass by a gravitational interaction.  One can 
show \cite{DeFrGh90s} that the Higgs field as source of 
the inertial mass of the elementary particles mediates a scalar
gravitational interaction of Yukawa type between those
particles that become massive as a consequence of the spontaneous 
symmetry breaking.  Due to the equivalence principle, it seems natural 
to identify both approaches.  
For this reason,  \cite{DeFrGh92,CeDe95ab} 
proposed a scalar-tensor theory of gravity where the Higgs field of 
elementary particles also plays simultaneously the role of a variable 
gravitational constant, instead of the scalar field introduced by Brans 
and Dicke. 

Put another way, if there is an interaction between a scalar (Higgs) field 
and matter, and of course there exists coupling between matter and 
gravity, why not close the loop by incorporating a (non-minimal) coupling 
between the scalar field and gravity?  Or conversely, if one explores 
scalar-tensor theories, and gravity couples with matter, why not include 
an explicit interaction between the scalar field and matter? 

In this work we employ these ideas merely to motivate the couplings.  
For example, one might try to identify the dark energy (DE) with a 
Higgs-type (but not {\it the\/} Higgs) field that is coupled 
to some dark fermion sector. Accordingly, in our model, gravity, i.e.\ the 
Ricci scalar $R$, couples to a scalar field $\phi$ through  the non-minimal 
coupling $\phi^{2} R$.  As the field evolves to its energy minimum the 
Higgs coupling might give rise to the mass of some dark fermion that would 
account for the dark matter (DM) of 
the model.  The scalar field evolves to a constant, to generate the mass,  
and simultaneously generates Newton's constant through
the non-minimal coupling to gravity.
The resulting theory is an induced gravity in which GR is 
dynamically obtained through a Higgs mechanism from a scalar-tensor 
theory \cite{DeFrGh92,CeDe95ab}.

The proposed Higgs mechanism is at present hypothetical, but  
phenomenologically interesting since it can account for the mass of the DM 
and since its field can act as a DE to accelerate the 
cosmological expansion.   
While many DM-DE interaction models aiming to unify the two quantities 
have difficulties in getting them simultaneously to exist and match 
observations, this can sometimes be made easier by adding a third 
element, such as inflation \cite{LiUr06,unifica06} or, as in the present 
work, gravity.  In any case, apart from the motivation, exploration 
of the dynamics of the matter- and gravity coupled system is of interest. 

We begin by describing the general field equations in Sec.~\ref{fe}, 
identifying the contributions of the gravity-scalar and scalar-matter 
couplings.  This is then examined for the homogeneous and isotropic 
FRW universe background in Sec.~\ref{frw}, including the evolution 
equations for the scalar field and matter.  Section~\ref{sym} 
discusses the effective potential, illustrating it for a 
symmetry breaking form.  In Sec.~\ref{evo} we solve for the cosmic 
evolution of the field and matter, for general classes of potential 
and coupling, leading to constraints on the allowed parameter space from 
the cosmic expansion behavior.

\section{Field equations \label{fe}}

The scalar-tensor theory Lagrangian used here is similar to the one 
studied in the past for 
inflationary dynamics \cite{CeDe95ab}:  
\begin{equation}\label{lag41}
\mathcal{L} = \frac{\alpha}{16\pi}\phi^{2} R  +
\frac{1}{2}\phi_{;\mu}\phi^{;\mu} - V(\phi) +
\mathcal{L}_M \, ,
\end{equation} 
where $\phi$ is a real scalar field and $\alpha$ is a dimensionless
parameter.  We use the metric signature $(+ - - -)$.
Rewriting the Lagrangian as 
\beqa \label{lagg} 
\mathcal{L}&=& \frac{1}{16\pi G_N} R + \frac{1}{16\pi G_N} 
\left(\frac{\phi^2}{v_{GR}^2}-1\right)\,R \nonumber  \\ 
&&+\frac{1}{2}\phi_{;\mu}\phi^{;\mu}
-V(\phi)+\mathcal{L}_M, 
\eeqa
one sees that, formally, the Einstein-Hilbert action with the standard 
Newton's constant $G_N$ corresponds to 
$\phi^2 \to v_{GR}^2 \equiv 1/(\alpha G_N)$. Note however that
even if on average $\phi = v_{GR}$, the theory will
still be distinct from GR because of perturbations in the field. 

Varying Eq.~(\ref{lag41}), one obtains the field equations 
\begin{eqnarray} \label{ecgragi}
R_{\mu \nu} - \frac{1}{2} R g_{\mu \nu} 
& =& - \frac{8 \pi}{\alpha \phi^{2}} \left[ T_{\mu \nu} +  
 V(\phi)  g_{\mu \nu} \right] 
  \nonumber \\ &  &
 - \frac{8 \pi}{\alpha \phi^{2}}
 \left[ \phi_{ ; \mu } \phi_{ ; \nu}
 - \frac{1}{2} \phi_{; \lambda} \phi^{; \lambda} g_{\mu \nu}\right]
 \nonumber \\
&  & - \frac{1}{\phi^{2}} \left[(\phi^{2})_{;   \mu ;  \nu} - 
{(\phi^{2})^{;  \lambda}}_{; \lambda} g_{\mu \nu} \right] \, ,
\end{eqnarray}
where a semicolon stands for a covariant derivative and 
$ T_{\mu \nu} \equiv  \frac{2}{\sqrt{-g}} \frac{\partial \sqrt{-g} 
\mathcal{L}_{M}}{\partial g^{\mu \nu}}$.  For 
the dark matter (DM) Lagrangian we begin with a general ``Yukawa" 
coupling, $\mathcal{L}_M = \bar{\psi} \left(\imag \gamma^{\mu} \pa_{\mu} - f(\phi) \right) \psi$,
similar to 
in \cite{DaCoKh06,Co08}, and later explore some particular forms.  
The dark matter thus has a $\phi$ dependent mass $f(\phi)$.

The scalar field fulfills the generalized Klein-Gordon equation 
\begin{equation} \label{sfeq}
{\phi^{;  \lambda}}_{;  \lambda} + \frac{\partial
V}{\partial \phi} - \frac{\alpha}{8 \pi}R \phi=  \frac{\partial
\mathcal{L}_{M}}{\partial \phi}\,.
\end{equation} 
Note the term involving $\alpha R$ arises from the non-minimal coupling 
to gravity, and the right hand side comes from the coupling to dark matter. 

Taking the trace of Eq.~(\ref{ecgragi}) and substituting it into 
Eq.~(\ref{sfeq}) to remove $R$, one obtains
\begin{equation} \label{sfeq2}
{\phi^{2 ; \lambda}}_{; \lambda} + \frac{2}{1+ \frac{3 \alpha}{4 \pi}} 
\left(  \phi \frac{\partial V}{\partial \phi} - 4  V(\phi)    
- T - \phi \frac{\partial \mathcal{L}_{M}}{\partial \phi} \right)  = 0 .  
\end{equation}
The first term in the parentheses is the normal GR term for an uncoupled 
scalar field (although with an altered prefactor),  the $T+4V$ of the second 
and third terms stem from the nonminimal coupling to gravity (specifically 
the trace of the first line of Eq.~(\ref{ecgragi}), and the last term is 
due to the DM-DE interaction.  

This equation  can be recast as:
\begin{equation} \label{sfeq3}
\frac{1}{2 \phi} \, {\phi^{2 ; \lambda}}_{; \lambda} + V_{\rm eff}' (\phi)  
\,  =  0, 
\end{equation}
where a prime stands for partial derivative with respect to $\phi$.  
The effective potential is 
\begin{equation} \label{veff1}
V_{\rm eff}(\phi) \equiv   
\frac{1}{1+\frac{3 \alpha}{4 \pi}}  \left[ V(\phi)-\mathcal{L}_{M}(\phi) - \int^\phi 
d\varphi \frac{4 V(\varphi) + T(\varphi)}{\varphi}   \right] ,  
\end{equation} 
from which the effective mass of the Higgs particle will be identified 
later on.   We write the argument $\varphi$ explicitly for $\mathcal{L}_M$ 
to remind that only the partial derivative with respect to the field is 
relevant.  
If $V(\phi) = V_0 \phi^4$, i.e.\ there is no intrinsic mass, 
then one sees that the potential terms cancel out and the effective mass 
is completely determined by the trace $T$ from the nonminimal gravity 
coupling and the DM-DE interaction term $\mathcal{L}_M$.  

It is convenient to define the energy momentum tensors associated with 
the different contributions to the right hand side of Eq.~(\ref{ecgragi}): 
\begin{eqnarray} \label{tmunu} 
T^\mu{}_\nu{}^{(\phi)} &\equiv& V(\phi) \delta^{\mu}{}_{\nu} + 
\phi^{; \mu} \phi_{; \nu}  -   
\frac{1}{2} \phi^{;\lambda} \phi_{;\lambda} \delta^{\mu}{}_{\nu} \\ 
T^\mu{}_\nu{}^{(R \phi)} & \equiv& 
\frac{\alpha}{8 \pi} \left[ (\phi^{2})_{; \nu ; \lambda} g^{\lambda \mu} - 
{(\phi^{2})^{; \lambda}}_{; \lambda} \delta^{\mu}{}_{\nu} \right] ,
\end{eqnarray}
where the first is the standard scalar field contribution and the second 
stems from the nonminimal coupling to gravity.  Again, note that the GR 
limit in which $T^{\mu}{}_\nu{}^{(R\phi)}$ vanishes is not $\alpha=0$ but 
$\phi\to\,$const.  Due to the couplings, the components of the 
energy momentum tensor are not individually conserved but the total 
energy momentum tensor is: 
\begin{equation} \label{conserv}
\left\{ - \frac{8 \pi}{\alpha \phi^2} \left[T^{\mu}{}_{\nu}  + 
T^{\mu}{}_\nu{}^{(\phi)} + 
T^{\mu}{}_\nu{}^{(R \phi)} \right] \right\}_{; \mu} = 0.
\end{equation}
The conservation equation for matter can be derived from this (using the 
equation of motion (\ref{sfeq}) for $\phi$ 
and the Einstein equations (\ref{ecgragi})), giving
\beq \label{conservmat}
T^{\mu \nu}{}_{;\nu}=-\phi^{;\mu}\,\frac{\pa \mathcal{L}_M}{\pa \phi}\,. 
\eeq 
We will use this equation in the following to determine the evolution of 
the matter component.

\section{FRW cosmology equations \label{frw}}

To study the background evolution of the universe we use the FRW metric
\beq
ds^2 = dt^2 - a^2(t) \left( \frac{d r^2}{1 - k r^2} + r^2 d\Omega^2\right),
\eeq
and assume 
that the DM fluid is given by $ T_{\mu \nu} = \rho \, u_{\mu} u_{\nu}$ 
with the energy density $\rho = n f(\phi)$, where $n$ is the number density,
and $u_\mu \equiv d x_{\mu}/ds = \delta^{0}_{\mu}$ the comoving 4-velocity.   
The energy momentum tensor behaves as a pressureless dust perfect fluid
because we assume the dark matter to consist of non-relativistic
fermions. The equivalence between the description in terms of the spinor field lagrangian
as specified in the previous section on the one hand,
and the ``dust'' description on the other hand is
demonstrated in \cite{FaPe04} for a standard linear
coupling ($f \propto \phi$). There, it is shown how  a fermion field interacting 
with a scalar field $\phi$, with action 
\begin{equation}
 S=\int d^4 \sqrt{-g} (i \bar{\psi}\gamma^{\mu}\partial_{\mu} \psi - y (\phi - \phi_* ) \bar{\psi}\psi),
\end{equation}
where $y$ and $\phi_*$ are constant, is equivalent to a model of a classical gas of pointlike particles,  with action  
\begin{equation}
 S= -\sum_i \int  y (\phi - \phi_* ) ds_i,
\end{equation}
in the limiting situation where the fermions' de Broglie wavelengths are much smaller than the
characteristic length scale 
of variation of the $\phi$ field.  Following the steps of that demonstration one can see that it is
valid to replace the factor $y (\phi - \phi_* )$ by an arbitrary function of  $\phi$, see also Ref. \cite{DaPo94}. In our case we start with a fermion field 
with a coupling $f(\phi) \bar{\psi}\psi$.  Then, a valid effective action is 
\begin{equation}
 S= -\sum_i \int  f(\phi) ds_i = - \int d^4 x \, \sqrt{-g} \, f(\phi) n(x),
\end{equation}
where
\begin{equation}
 n = \sum_i \int ds_i \frac{\delta^4 (x - x(s_i))}{\sqrt{-g}} .
\end{equation}
This is exactly the action of a dust fluid with energy density $\rho = f(\phi) n$.

Thus, the matter Lagrangian is proportional to $f(\phi)$,
i.e. $\mathcal{L}_M = - T = - \rho \propto f(\phi)$.  
From the gravity field Eq.~(\ref{ecgragi}) one obtains the generalized 
Friedmann equations 
\begin{equation}\label{2da} 
\frac{\dot{a}^2 + k}{a^2} =  \frac{8 \pi}{3 \alpha \phi^{2}} 
\left( \rho + V(\phi) + \frac{1}{2}   \dot{\phi}^2   
-  \frac{3 \alpha}{4 \pi} H \phi \dot{\phi} \right) , 
\end{equation}
\begin{eqnarray}\label{acc2}
2\frac{\ddot{a}}{a} + \frac{\dot{a}^2+k}{a^2} &=&
\frac{8\pi}{\alpha \phi^2} \left( V(\phi) - \frac{1}{2} \dot{\phi}^2 \right)  \nonumber \\  
&&   - \frac{2}{\phi}  \left(  \ddot{\phi} + 2 H \dot{\phi}  + \frac{\dot{\phi}^2}{\phi} \right),
\end{eqnarray}
where $H \equiv \dot{a}/a$ is the Hubble parameter.

Rearranging the terms, the acceleration equation is 
\begin{eqnarray}\label{acc3}
\frac{\ddot{a}}{a} &=&  - \frac{4 \pi}{3\alpha \phi^{2}} 
 \left[\rho - 2 V(\phi) \right] \nonumber\\  
&\,& - \left( 1 + \frac{8 \pi}{3 \alpha} \right) \frac{\dot{\phi}^2}{\phi^2}  
- H \frac{\dot{\phi}}{\phi} - \frac{\ddot{\phi}}{\phi}  \,.
\end{eqnarray}
The redundant Klein-Gordon equation~(\ref{sfeq3}) becomes 
\begin{equation} \label{sdcampo}
\ddot{\phi} + \frac{\dot{\phi}^2}{\phi} + 3 H\dot{\phi} + V_{\rm eff}{'} = 0\,,
\end{equation} 
where 
\begin{equation}
V_{\rm eff}(\phi) = \frac{1}{1+\frac{3 \alpha}{4 \pi}} \left[ V(\phi) 
+ \rho(\phi)  - \int^\phi d\varphi \frac{4 V(\varphi) + \rho(\varphi)}{\varphi} \right] . \label{veff2}
\end{equation}
Since the field is coupled to matter (either employing a Higgs or other 
mechanism), the source term $V_{\rm eff}{'}$ involves the density 
of the interacting DM fluid as well as the potential of the DE field. 

The conservation equation, Eq.~(\ref{conservmat}), yields 
\begin{equation} \label{eceq1}
\dot{\rho}+3 H \rho = - \mathcal{L}_M{'} \dot{\phi} = \dot{\phi} \, \rho \frac{f^{'}}{f} \,, 
\end{equation} 
so that the matter behaves the same way as in GR with a DM-DE 
interaction \cite{DaCoKh06}.  This equation can be directly integrated 
to give 
\begin{eqnarray}  \label{eceq-sol}
\rho= \frac{n_{0}}{a^3} f(\phi) ,
\end{eqnarray}
where $n_0$ is the DM number density at present.
This simply tells us that DM particle number is conserved and 
the change in energy per comoving volume is purely due to the 
varying mass $f(\phi)$.

\section{Effective Potential \label{sym}}

The nonminimal coupling to gravity, and the coupling to matter, in 
addition to adding symmetry to the relations between the scalar 
field, matter, and gravity, also can create a nonzero vacuum expectation 
value (vev) -- an effective cosmological constant -- that can adiabatically evolve.  
The nonminimal gravity coupling can also reduce the driving term 
$V_{\rm eff}{}'$ (this coupling gives rise to the negative term in 
Eq.~\ref{veff2}), slowing the field down.  This slow roll can often 
alleviate instabilities in coupled matter perturbations 
\cite{Co08}. 

We can examine these influences in terms of the effective potential 
of the theory, which alters the bare scalar field potential through 
the coupling to dark matter and to gravity.  
Using the Yukawa coupling $f  \propto \phi$ in Eq.~(\ref{veff2}), the effective 
potential becomes 
\begin{equation} \label{veffp}
V_{\rm eff}(\phi)=\frac{1}{1+\frac{3\alpha}{4\pi}} 
\left[V(\phi)-4\int^\phi d\varphi\,\frac{V(\varphi)}{\varphi}\right]\,. 
\end{equation}
Note that the density terms in Eq.~(\ref{veff2}) -- the 
interacting DM term $ \mathcal{L}_{M}^{'} $ and the trace term  $T$    
stemming from the nonminimal coupling $\phi^2 R$ -- cancel out 
since $\rho$ is linearly proportional to $\phi$.   If  
furthermore $V(\phi) = V_0 \phi^4$, then the effective potential 
vanishes and the dark energy field, even coupled, is massless and acquires 
infinite range. This is because in this case the theory has no explicit 
mass scale in it and is thus scale invariant.

For the Yukawa coupling, the effective potential does not run with the 
density, so it differs from -- and is actually simpler than -- what 
happens in the GR case with matter coupling.  The effective mass is only 
determined by the potential terms.  The field acts like quintessence with a 
$V_{\rm eff}(\phi)$ given by the full Eq.~(\ref{veffp}). 

For a Landau-Ginzburg symmetry breaking form for the Higgs potential, 
\begin{equation} \label{langin} 
V(\phi) = \frac{\lambda_a}{24}\left(\phi^{2} - 
\frac{6\mu^2}{\lambda_a}\right)^2  \,,  
\end{equation} 
where $\lambda_a$ is a dimensionless constant, $\mu\sqrt{2}$ is
the mass of the field at the potential minimum, 
and the Higgs ground state $v$, such that $V(v)=0$, is given by 
\begin{equation}
v^2 = \frac{6\mu^2}{\lambda_a} \,. 
\end{equation}

In induced gravity, the Higgs potential $V(\phi)$ generates a time 
varying gravitational coupling (as in scalar-tensor theory) 
\begin{equation} \label{gfsm} 
G(\phi)= \frac{1}{\alpha  \phi^2}
\end{equation} 
as $\phi$ rolls from an initial state to its 
ground state and thus for a given field value determined by the potential,
$\alpha$ needs to be chosen such that $G(\phi) \approx G_N$.  
For example, we might choose $\alpha$ such that $G(\phi_{\rm min})=G_N$. 

Relating this to the particle physics of the Higgs mechanism, one has 
\begin{equation}
\alpha = 2\pi \left(g \frac{M_{Pl}}{M_b}\right)^2  \, ,
\end{equation}
where $M_{Pl} \equiv 1/\sqrt{G_N} \approx 1.2 \times 10^{19}\,$GeV is the 
Planck mass, $M_b$ the boson mass, and $g$ a coupling constant.  
If we were to consider the standard model Higgs, one has 
$M_b= M_{\mbox{w}} = 80\,$GeV for the W-boson and $g = 0.18$, therefore
$\alpha \approx 10^{33}$.  Such a value is huge and would not pass
cosmological constraints on $\alpha$ discussed 
below. Therefore, we consider another Higgs-like 
particle with a much larger mass than that of the Higgs of the 
standard model of particle physics.  The parameter $\alpha$ 
will need to be determined through cosmological observations. 

The effective potential for the Landau-Ginzburg plus Yukawa coupling case is 
\begin{equation} \label{veff}
V_{\rm eff}  = \frac{1}{1+\frac{3 \alpha}{4 \pi}} \frac{\lambda_a v^4}{12} 
 \left[ (\phi^2/v^2) - {\rm ln} (\phi^2/v^2) \right] \,, 
\end{equation} 
and has a minimum at $\phi =v$ at all times, at which the mass of Higgs 
particle is 
$m_H^{2} = V_{\rm eff}{}''(\phi =v)=\lambda_a v^2/(3[1+3\alpha/(4\pi)])$. 

\begin{figure}[!htbp] 
\includegraphics[width=\columnwidth]{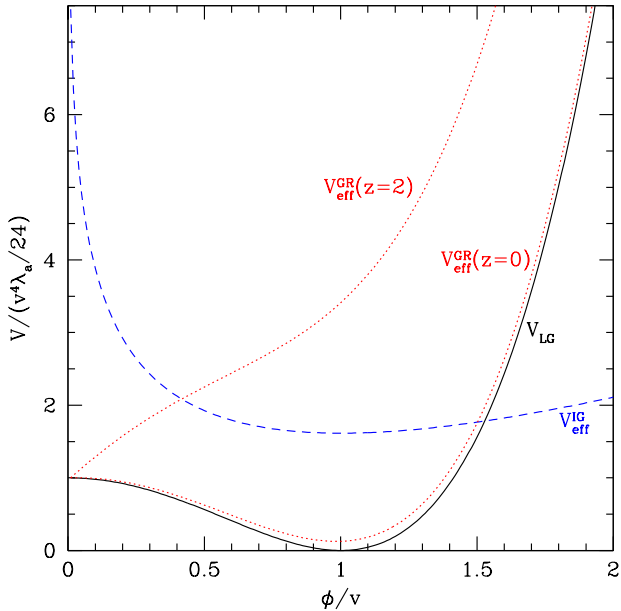}  
\caption{The Landau-Ginzburg potential in the uncoupled, GR case (solid 
curve), in the matter coupled but minimally gravity coupled (GR) case 
for two values of the redshift $z$ (dotted lines), and in the nonminimally 
gravity coupled induced gravity (IG) case (dashed curve).  The matter 
coupling is taken to be of the Yukawa form.} 
\label{fig:vefflg} 
\end{figure}

We show the difference between the bare potential and the effective 
potentials in induced gravity and in GR with Yukawa coupling in 
Fig.~\ref{fig:vefflg}. 
Since the matter density contribution cancels out in induced gravity, 
the effective potential is time independent, while the GR case is not. 
The induced gravity case gives a broad, nearly flat minimum. 
When the field is very slowly rolling, the modifications to the right 
hand side of Eq.~(\ref{ecgragi}) are small.

In this Ansatz the Higgs field, rolling along the effective 
potential, plays the role of DE and the particle masses produced by the Higgs 
mechanism provide the DM of the model.  However, as we will see the 
dynamics does not generally favor such a form for the potential.

\section{Dynamical Equations} \label{evo}

We now examine the dynamics of the dark energy, the effective 
equation of state of the dark matter, and the overall expansion behavior. 

\subsection{Variables and Equations of Motion} 

The dynamics of the scalar field, non-minimally coupled to gravity and 
to dark matter, are described by an autonomous system of equations 
which can be solved in a straightforward manner.
One can define the 
following set of variables \cite{CoSaTs06}:
\begin{eqnarray} \label{var-def}
x &\equiv& \frac{\kappa \dot{\phi}}{\sqrt{6}H} \qquad ;\qquad 
y \equiv \frac{\kappa \sqrt{V}}{\sqrt{3}H} \nonumber \\
\lambda &\equiv& -\frac{V^{'} }{\kappa V}= 
- \sqrt{\frac{\beta}{6}} \, \frac{d\ln V}{d\ln\phi}  \quad ;\quad 
N \equiv  \ln a \,, 
\end{eqnarray}
where $\kappa=\sqrt{6/\beta}/\phi$ is a function of the field, with 
$\beta \equiv 3\alpha/(4 \pi )$.  

Assuming spatial flatness, i.e.  $k = 0$, the autonomous 
system becomes 
\begin{eqnarray} \label{auton-sys}
\frac{d x}{d N} &=&
- \frac{3}{2} x y^{2}-\frac{3}{2}x-3\sqrt{\beta} \, x^{2} + \frac{3}{2} x^{3} 
 \nonumber \\
& & 
+ \frac{1+\sqrt{\beta}\,x}{1+\beta} 
\left(  \sqrt{\frac{3}{2}}  \lambda y^{2} -  C \right) \,, \\
\frac{d y}{d N} &=&
y \left[- \sqrt{\frac{3}{2}} \lambda x  + \frac{3}{2} \left(  x^{2} - y^{2} +1 \right) 
- 2 \sqrt{\beta}  \, x \right. \nonumber \\
&\qquad & \left. + \frac{\sqrt{\beta} }{1+\beta}\,
 \left(\sqrt{\frac{3}{2}}\lambda y^{2} - C\right) \right] \,. 
\end{eqnarray} 
Here 
\begin{eqnarray}  \label{termc1}
C &\equiv& \frac{\kappa}{\sqrt{6}H^2}\left(\rho\,\frac{f'}{f}-
\frac{4V+\rho}{\phi} \right) \\ 
&=&\frac{\sqrt{\beta}}{2}\, 
\left[\left(F-1 \right)\,\Omega_m(N)-4y^2\right]\,, 
 \label{termc2}
\end{eqnarray} 
where $F\equiv d\ln f/d\ln\phi$.   The quantity $C$ encodes the 
information on the matter coupling (in the first term involving $f$) and 
the non-minimal gravitational  coupling.  

The fractional matter density comes directly from Eq.~(\ref{2da}) as 
\begin{eqnarray}
\label{constraint}
\Omega_m(N)&\equiv& \frac{\kappa^2\rho}{3H^2}=1- \Omega_{\phi} \nonumber \\  
\Omega_{\phi} &=&  x^2 + y^2 - 2\sqrt{\beta} \, x \,. 
\end{eqnarray}  
To evolve the field $\phi$ appearing in $\lambda$ or in $d\ln f/d\ln\phi$, 
one uses the auxiliary equation 
\begin{equation}
\frac{d\ln\phi}{dN}=\sqrt{\beta}\,x \,. 
\end{equation} 

We define the effective equation of state for a component in terms of 
$\kappa^2\rho_i$ (since the sum of these quantities is conserved, 
see Eq.~\ref{conserv}), by
\beq 
w_i=-1-\frac{1}{3}\frac{d\ln\kappa^2\rho_i}{dN} \,. \label{eq:wdef} 
\eeq
This takes into account effective pressure terms due to the interactions. 
In particular, the matter equation of state gains an effective nonzero term, due to the 
nonminimal coupling to both the scalar field directly and to gravity 
through the $\phi^2 R$ term.  One has
\beq 
w_m=\frac{\sqrt{\beta}}{3}\,(2-F)\,x \, .
\eeq 
One has $w_m=0$ only when the field is frozen ($x=0$), which restores GR 
as discussed in Sec.~\ref{fe}, or when $f\sim\phi^2$,  
which counteracts the $\kappa^2$ term in Eq.~(\ref{eq:wdef}). 

The equation of state for the scalar field is more involved: 
\beqa \label{omega_phi} 
w_{\phi} &=& \left[y^{2} [\beta  (F-U)-3]-(F-2) x^{3} \sqrt{\beta } (\beta +1) \right.  \nonumber \\
&&\left. +x^{2} (\beta [2(F-2)\beta +3 F-2]+3) -\beta(F-1)\right. \nonumber \\
 && \left. -x \sqrt{\beta}   \left[y^2(F-2)(\beta+1)+\beta(2+F)+4-F\right] \right] \nonumber\\ 
&& \big / \left[ 3 (\beta +1)  \left(x^{2}-2 x \sqrt{\beta}+y^{2} \right) \right] \, ,
\eeqa
which depends on the nonminimal gravity coupling through $\beta$, on the 
interacting DM-DE term ($F$), and on the potential 
($U\equiv d\ln V/d\ln\phi$). 

The total equation of state 
\beqa 
w_{\rm tot}&\equiv& -1 -\frac{1}{3}\frac{d\ln H^2}{dN} \\ 
&=& w_m(a)\,\Omega_m(a) + w(a)\,\Omega_\phi(a) 
\eeqa 
takes the same form with or without coupling to matter.  The 
deceleration parameter $q=(1+3w_{\rm tot})/2$ as usual.

\subsection{Critical points}

Investigating the critical (or fixed) points of the dynamical system, 
we find the 
relation (for $y_c\ne0$) 
\beq 
y_c^2=1+\sqrt{\beta}\,\,\frac{U+2}{3}\,x_c+
\left(1+\beta\,\frac{U+2}{3}\right)x_c^2 \,.  
\eeq 
We restrict to the cases of $U$ and $F$ being constant in this 
analysis; otherwise the equations are transcendental. 
The dark energy density and equation of state are 
\beqa 
\Omega_{\phi,c}&=&1-\frac{\sqrt{\beta}}{3}(4-U)\,x_c 
+\frac{1}{3}\left[6+\beta(2+U)\right] x_c^2 \\ 
w_{\phi,c}&=&-1+\frac{\sqrt{\beta}}{3}(2-U)\,x_c\,. 
\eeqa 

\begin{figure}[!htbp]
\includegraphics[width=\columnwidth]{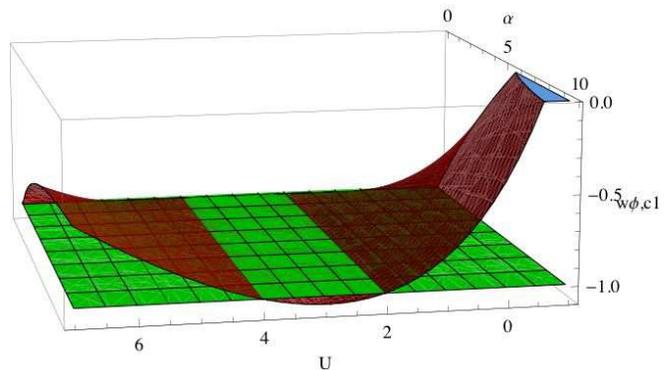}
\caption{The dark energy equation of state for the first critical point 
is plotted as a function of the power law index $U$ of the potential 
and the value $\alpha$ of the gravity coupling (the matter coupling 
does not affect this critical point).  The value $w=-1$ is indicated by the 
green mesh plane.  Note the attractor gives acceleration, and $w\approx-1$, 
over a wide range of values.} 
\label{wphic1}
\end{figure}

The fixed point $x_c$ is given by a quadratic equation.  The first 
solution is 
\beqa 
x_{c1}&=&\sqrt{\beta}\,\,\frac{4-U}{6+\beta(2+U)} \\ 
\Omega_{\phi,c1}&=& 1 \\ 
1+w_{\phi,c1}&=&\frac{\beta}{3}\frac{(2-U)(4-U)}{6+\beta(2+U)} \,. \label{eq:wc1} 
\eeqa 
This is a stable fixed point over a wide range of parameters (see 
next section) and it is independent of the matter coupling 
$f(\phi)$.  Note that we obtain a future de Sitter state ($w=-1$) 
for $V\sim\phi^2$ or $V\sim\phi^4$ asymptotic behavior.  
The $U=4$ case represents an attractor to GR, since in this case 
$x_{c} =0=\dot{\phi}$.  An $x_c=0$ solution was also found in GR with 
no matter couplings \cite{GRattractor,DaPo94,BaPi00}, and with an exponential 
coupling \cite{BeFlLaTr08}, and also in induced gravity 
with no matter couplings \cite{FaJe06,FiTrVe08}.  
For $2<U<4$ the fixed point gives a phantom ($w_c<-1$) 
attractor; a phantom solution was obtained in Ref.~\cite{Pe05} for 
$F=0$ and a particular choice of $U$.  
The general attractor solution is illustrated in Fig.~\ref{wphic1}.  
Note that there is a wide 
range of $U$ and $\alpha$ (and all of $F$, which does not enter) where 
the attractor gives acceleration, and in fact stays close to $w=-1$. 
In all cases $\alpha \to 0$ makes $w_{\phi,c1}  \to -1$.

The second solution (mostly a saddle point) does depend on $f(\phi)$ and 
is given by 
\beqa
x_{c2}&=& \frac{3}{(F-U)\sqrt{\beta}} \label{eq:xc2} \\ 
\Omega_{\phi,c2}&=& \frac{18+\beta(6+7U-4F+F^2-FU)}{\beta (F-U)^2} \\ 
w_{\phi,c2}&=&\frac{2-F}{F-U}\,. 
\eeqa 
Again, $w_c=-1$ when $V\sim\phi^2$ (except when $F=2$ also, since for 
$F=U$ this solution does not exist).  A solution within GR for 
an exponential potential and an exponential coupling function \cite{Am00} 
also has a critical point depending on the equivalent of $F$ and $U$. 

Note that $x^2$, $y^2$, and 
$\Omega_m$ can all be greater than unity.  This is possible because 
of the negative term $-H\phi\dot\phi$ in the modified Friedmann 
equation (\ref{2da}), which permits the effective dark energy density 
to go negative.  This causes no physical problems within our Ansatz, 
as we demonstrate in Sec.~\ref{sec:dynhist}. 

Three further mathematical solutions exist for $y_c=0$.  These have 
$x_{c4,c5}=\sqrt{\beta}\pm\sqrt{1+\beta}$, which are mostly unstable 
and inaccessible, but of 
more interest is 
\beqa 
x_{c3}&=&\frac{\sqrt{\beta}\,(1-F)}{3+\beta(2+F)} \\ 
\Omega_{\phi,c3}&=&\frac{\beta(F-1)[5+F+2\beta(2+F)]}{[3+\beta(2+F)]^2}\\ 
w_{\phi,c3}&=&\frac{\beta(F-1)(F-2)}{3\,[3+\beta(2+F)]} \,. 
\eeqa 
This solution is important because $y\approx 0$ corresponds to 
$\kappa^2 V/(3H^2)\approx 0$, which holds at early times.  
We indeed find a metastable 
attractor to this behavior at high redshift (basically, 
$dx/dN\approx 0$ but $y$ keeps growing).  Note that this 
high redshift attractor is generally $w_\phi=w_m$, i.e.\ a scaling 
solution.  However, $V$ (and $y$) are not actually zero, just small 
compared to the other densities, so the case $F=1$, which would give $x_c=0$,  
breaks the condition $y_c\ll x_c$ and instead here the dynamics forces 
$w_\phi=-1$ and $w_m=0$.

\subsection{Stability} 

Carrying out a linear stability analysis, we find that the stability 
of the critical points generally depends on the values of $\alpha$ (i.e.\ 
$\beta$), $U$, and $F$, making analytic statements difficult.  However, a 
reasonable rule of thumb is that for values of these variables not too 
positive or too negative, the first critical point is a stable node (both 
eigenvalues of the perturbation matrix negative) and the second critical 
point tends to be a saddle point (one eigenvalue positive, one negative). 
See \cite{CoSaTs06} for a general discussion of stability analysis and 
classification. 

\begin{figure}[!htbp]
\includegraphics[width=0.95\columnwidth]{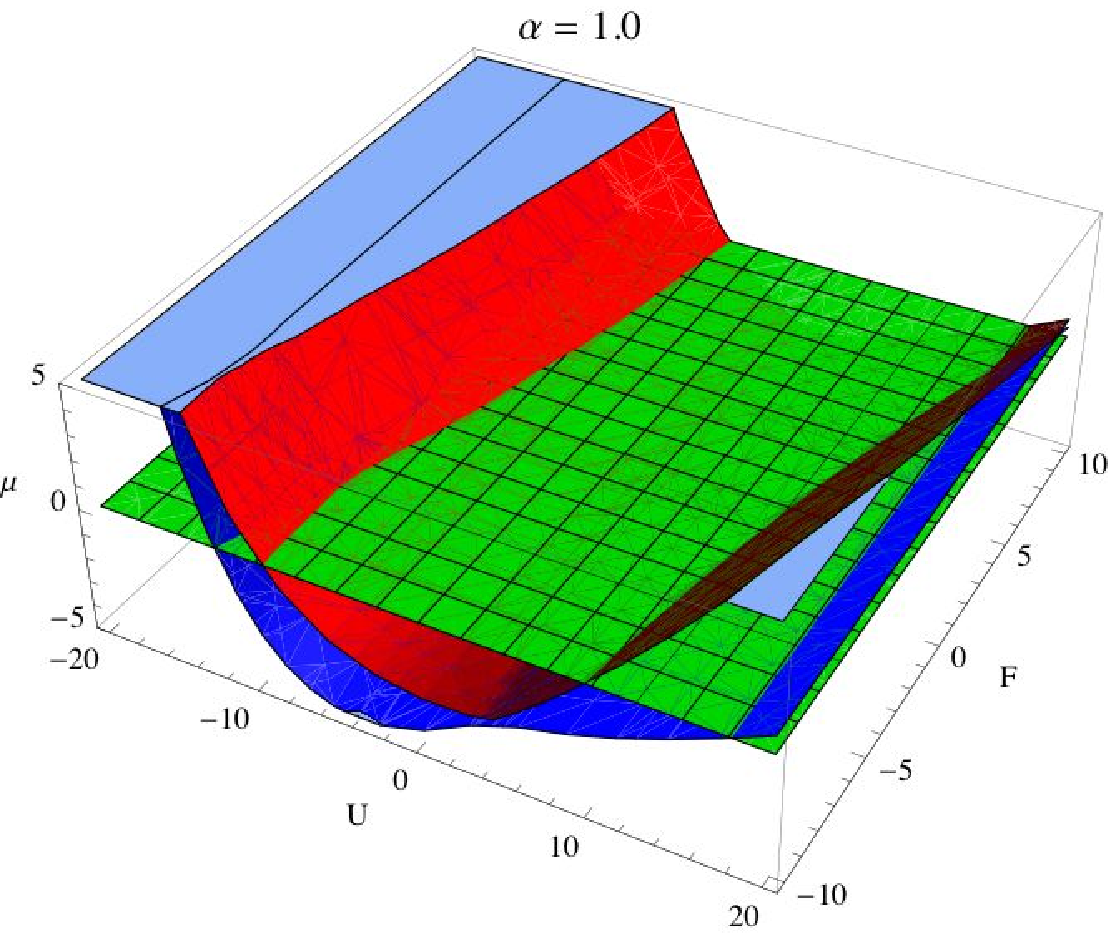}
\includegraphics[width=0.95\columnwidth]{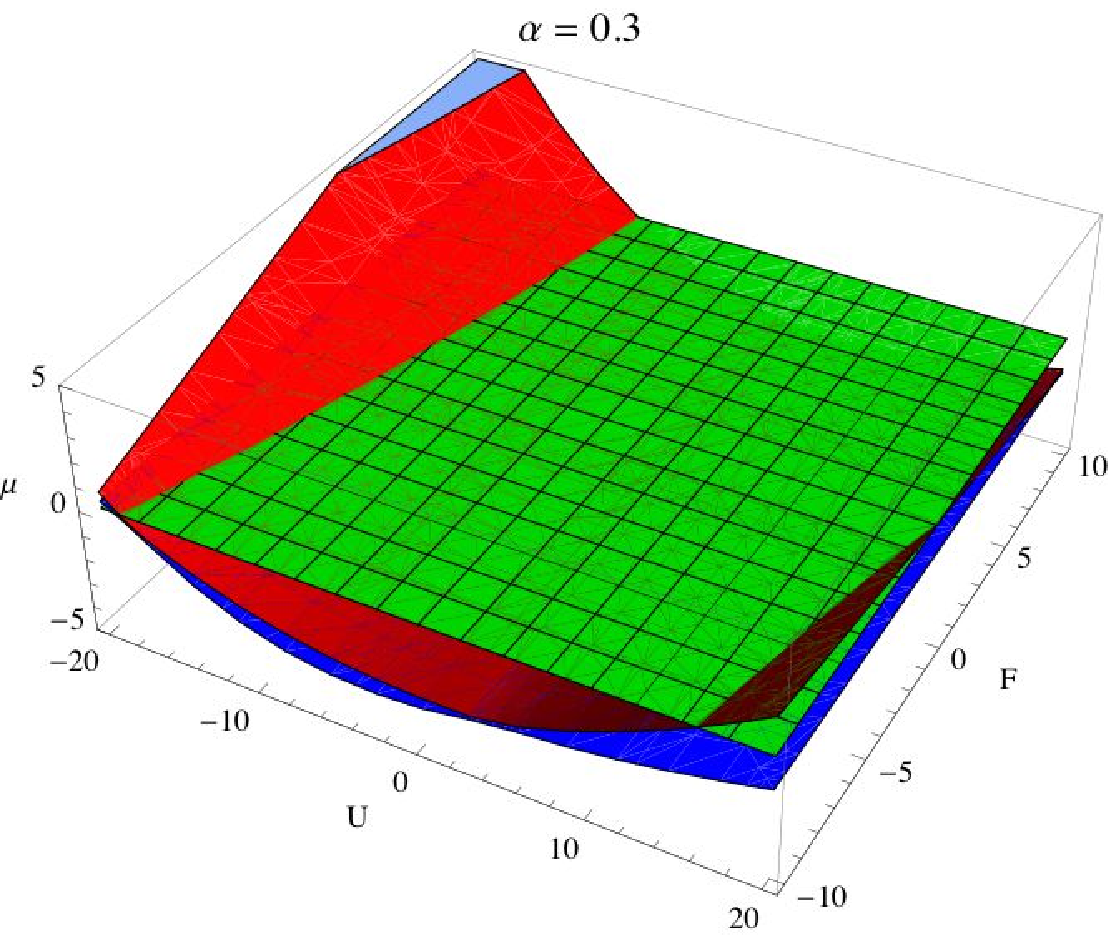}
\includegraphics[width=0.95\columnwidth]{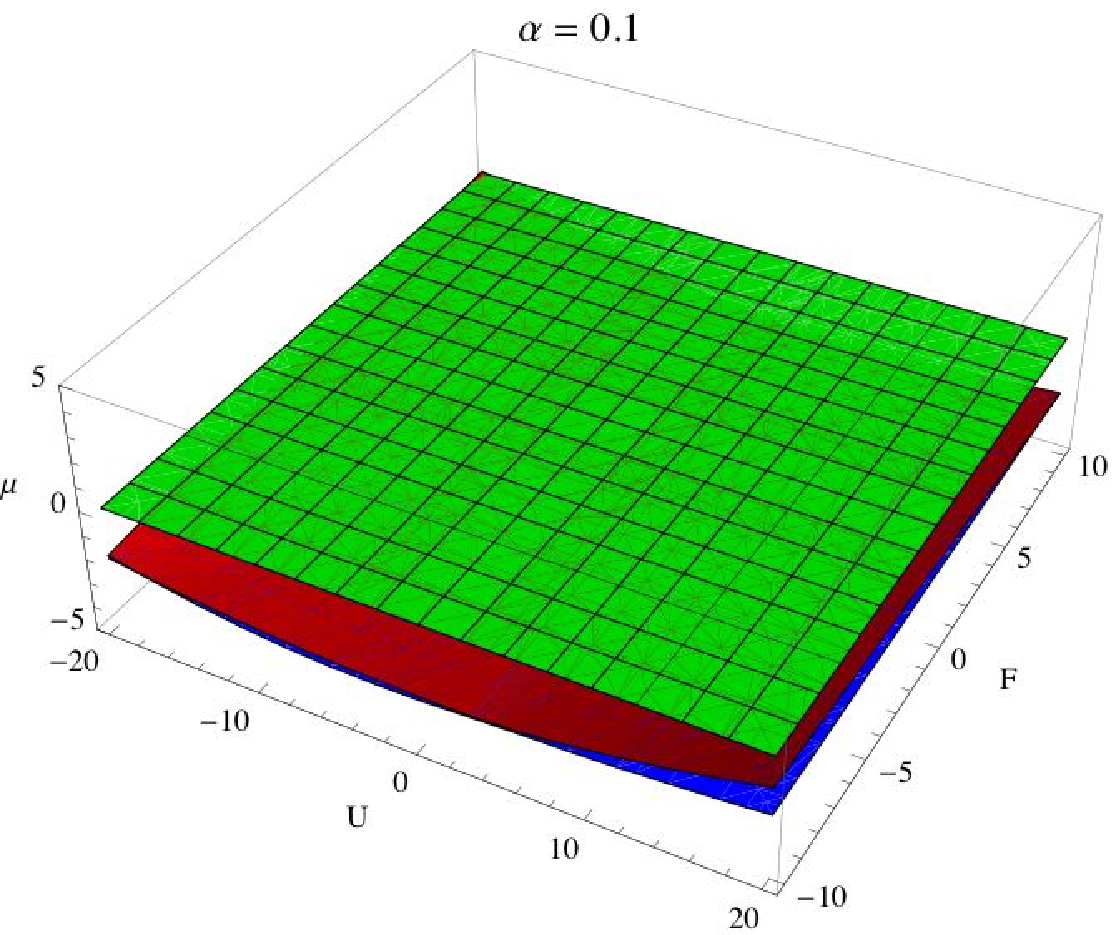}
\caption{Surfaces for the two eigenvalues $\mu_1$ (upper red) and $\mu_2$ 
(lower blue) of the first critical point.  The stability region for the 
critical point is where both eigenvalues are negative, i.e.\ below the 
(green horizontal) $\mu=0$ plane, and is a function of the power law index 
$U$ of the potential, $F$ of the matter coupling, and the value $\alpha$ 
of the gravity coupling.  The three panels shows how the surfaces change 
with $\alpha$: the smaller $\alpha$, the more stable the system.} 
\label{muc1}
\end{figure}

We show the 3-dimensional eigenvalue surfaces for the first critical 
point in Fig.~\ref{muc1}, for three different values for $\alpha$.   
The region of stability exists for a broad interval of $U$ and $F$, 
including ``natural'' values.  As $\alpha$ decreases, the area of 
stability grows, as seen in the sequence of the three plots.   
For $\alpha < 0.1$  the system is stable for any $U\in[-20, 20]$, and 
$F \in[-10, 10]$.

\begin{figure}[!htbp]
\includegraphics[width=\columnwidth]{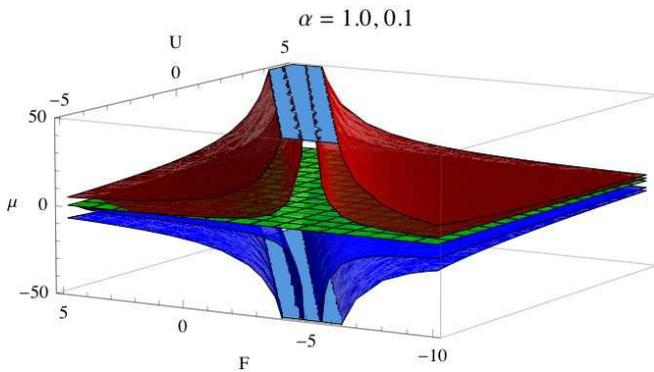}
\caption{Surfaces for the two eigenvalues $\mu_1$ (upper red) and $\mu_2$
(lower blue) of the second critical point.  Since one eigenvalue is 
positive, this critical point is not stable.  As $\alpha$ decreases, 
the sheets pull away from the zero plane.} 
\label{muc2}
\end{figure}

By contrast, the second critical point is a  saddle point,  
as seen in Fig.~\ref{muc2}.  One eigenvalue is always positive and the 
other negative, independent of the chosen $U$ and $F$.  As $\alpha$ 
decreases, the eigenvalue sheets tend to pull further away from 
the zero plane.  Since the qualitative 
nature of one positive eigenvalue, one negative eigenvalue does not 
change, we show two values of $\alpha$ together in the figure.  
Also note that the eigenvalues diverge for $F=U$, as expected from 
Eq.~(\ref{eq:xc2}).  

The third critical point is also a saddle point for 
$U\in[-10, 20]$ and  $F \in[-10, 10]$, but it has a stability window for 
$\alpha=1$ in the region $U\in[-10, -20]$ and  $F \in[-1, -5]$.  As 
$\alpha \to 0$ the sheets flatten, one below and above the $\mu=0$ plane, 
making this a saddle point for any $U$ and $F$.  Finally, the fourth and 
fifth fixed points are mostly unstable, with small regions of stability 
(for big $|F|$ and $|U|$) for $\alpha\ge 0.5$, but these solutions are 
anyway uninteresting since they are nonaccelerating and require $y=0=V$ 
while $\Omega_\phi=1$, hence $\phi=0$ and vanishing gravity.  

A key point is that $U=4$ always has a stable attractor to the first 
critical point, with $w=-1$ and $\dot\phi=0$, for all values of $\alpha$ and 
$F$ (even $F\ne$ constant, as we discuss in the next section).  So for a 
quartic potential there is an attractor leading to $\Lambda$CDM and a 
restoration of GR.

\subsection{Dynamical History} \label{sec:dynhist} 

\begin{figure}[!htbp]
\includegraphics[width=\columnwidth]{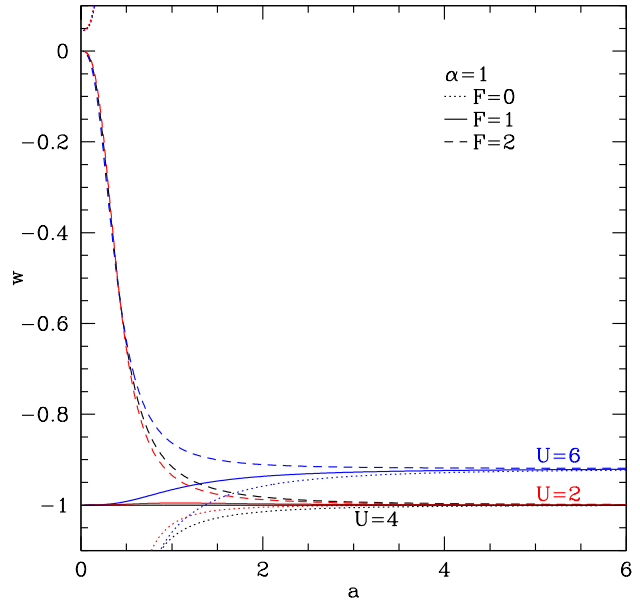}
\caption{The dark energy equation of state is plotted for three 
power law indices of the potential and the matter coupling.  All solutions 
asymptote to the stable fixed point $w_{c1}(U)$, independent of $F$.  Note 
both quadratic and quartic potentials ($U=2,\,4$) reach the 
de Sitter state $w=-1$. 
The deviations of $w(a)$ from $-1$ scale linearly with $\alpha$ for 
small $\alpha$.} 
\label{fig:dynw}
\end{figure}

Turning now to the full dynamics, some numerical solutions are 
illustrated in Fig.~\ref{fig:dynw}.  At high redshift, where 
$H^2$ is dominated by the matter density, $y$ is small and 
the field quickly forgets the initial conditions and goes to the 
scaling solution of the third critical point (except when $F=1$, then 
$w_\phi=-1$).  As the contribution of the dark energy potential energy 
becomes relatively more important, it then approaches the asymptotic 
attractor of the first critical point.  

We define the present ($a=1$) by when $\Omega_\phi=0.72$.  For clarity 
we show in Fig.~\ref{fig:dynw} the results for a value of the gravity 
coupling $\alpha=1$; for small $\alpha$ 
the deviation from $w(a)=-1$ will scale roughly as $\alpha$.   
The high redshift metastable attractor and the asymptotic future stable 
attractor behaviors can clearly be seen.  Note that for the 
Yukawa coupling ($F=1$) the dynamics stays close to $w=-1$.  For $F<1$, 
the $-H\phi\dot\phi$ term can drive the effective dark energy density 
through zero, making $w$ go to $\pm\infty$.  This has no physical 
pathology, since the dark energy density is merely an effective quantity 
and as we will soon see the matter and total equations of 
state are well behaved.  In the future, the attractor solutions for 
$V\sim\phi^2$ and $V\sim\phi^4$, i.e.\ $U=2$ and 4, are the de Sitter 
state $w=-1$.  The latter case represents the attractor to GR.

\begin{figure}[!htbp] 
\includegraphics[width=\columnwidth]{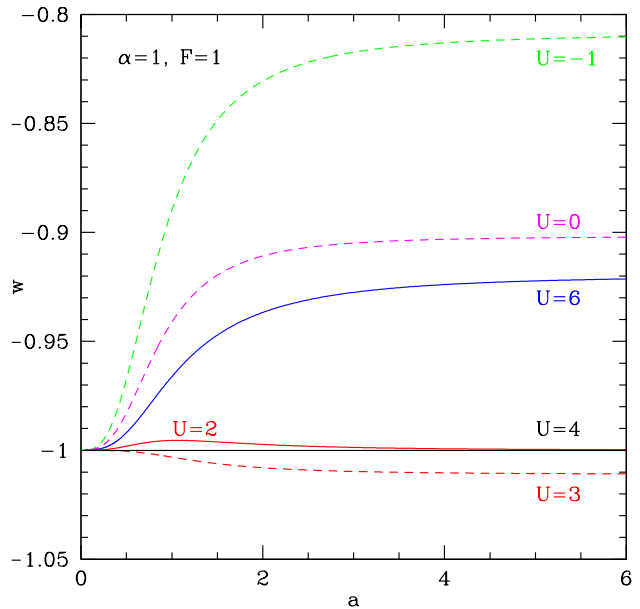}
\caption{The evolution of the dark energy equation of state is 
shown for various power law potentials $V\sim\phi^U$, with Yukawa 
coupling $f\sim\phi$ and gravity coupling $\alpha=1$.  The attractor 
values $1+w_c$ scale with $\alpha$ for small $\alpha$.  Note that 
$\phi^2$ and $\phi^4$ potentials give $w=-1$ attractors, while 
power law indices in between give phantom attractors. } 
\label{fig:pow}
\end{figure}

Examining the future attractor solution more closely, we see from 
Eq. (\ref{eq:wc1}) that 
within the range $2<U<4$, the attractor is to a phantom state $w<-1$, 
while outside this range it is to $w>-1$ (unless $U$ gets extremely 
negative). In general, reasonable inverse power law potentials do not 
provide $w\approx-1$ and would be disfavored by observations. 
Figure~\ref{fig:pow} shows the full behaviors for a variety of $U$ values, 
fixing the matter coupling to the Yukawa form $f\sim\phi$ ($F=1$).

\begin{figure}[!htbp] 
\includegraphics[width=\columnwidth]{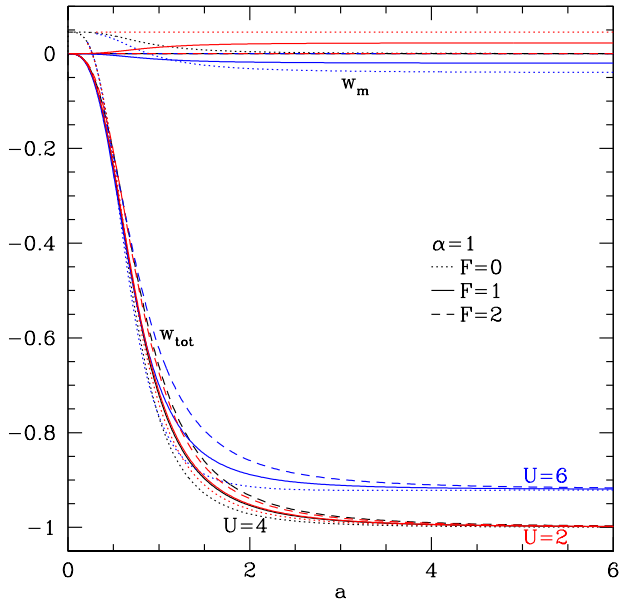}
\caption{The matter equation of state $w_m$ and total equation of state 
$w_{\rm tot}$ are plotted for the same potential and matter coupling cases 
as in Fig.~\ref{fig:dynw}.  The asymptotic future values of $w_m$ and 
$1+w_{\rm tot}$ scale linearly with $\alpha$ for small $\alpha$.  For Yukawa 
coupling ($F=1$), until recently $w_m$ was zero and $w_{\rm tot}$ 
followed the standard $\Lambda$CDM history (solid black line, also for 
$U=4$, $F=1$) as well.}
\label{fig:wmtot}
\end{figure}

Regarding the early time behavior, as stated the dark energy equation 
of state becomes undefined when the dark energy density passes through 
zero, as can occur for $F<1$.  Since the dark energy density vanishes, 
however, this has no physical effects.  In particular, the matter 
equation of state $w_m$ and total equation of state $w_{\rm tot}$ 
exhibit no sign of this occurrence.  In Fig.~\ref{fig:wmtot} we see that 
at high redshift $w_m$ goes to its metastable attractor solution (third 
critical point), and because of scaling $w_{\rm tot}=w_m$.  In the Yukawa 
coupling case ($F=1$), $w_m=0$, as it is for the $F=2$ case always as well.  
Both $w_m$ and $w_{\rm tot}$ then smoothly evolve toward the future stable 
attractor solution.  Both $w_m$ and $1+w_{\rm tot}$ scale linearly with 
$\alpha$ for small $\alpha$.  For $F=1$, all quantities are well behaved 
and follow $\Lambda$CDM until recently.

\begin{figure}[!htbp]
\includegraphics[width=\columnwidth]{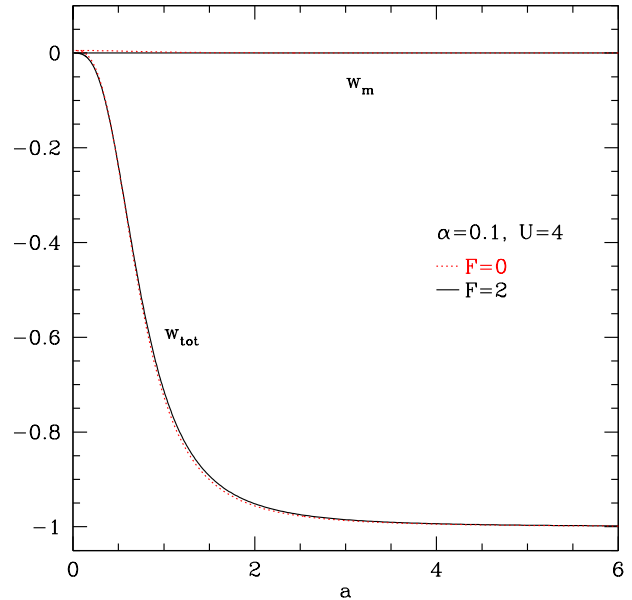}
\caption{As the $U=4$ case of Fig.~\ref{fig:wmtot}, but for $\alpha=0.1$. 
Deviations of $w_m$ from zero, and $w_{\rm tot}$ from the $\Lambda$CDM 
behavior are less than $10^{-2}$.} 
\label{fig:alf4}
\end{figure}

When $\alpha$ is small, the equations of state of matter, dark energy, 
and the total energy are nearly the same as for $\Lambda$CDM, regardless of 
$F$ and $U$.  This is demonstrated 
in Fig.~\ref{fig:alf4} for $\alpha=0.1$.

A small value of $\alpha$ corresponds to $\phi$ greater than 
the Planck energy to induce $G_N$ (see Eq.~\ref{gfsm}). 
However, there is no actual energy density that we are taking to be 
greater than the Planck scale, so it is not clear there is a problem 
in treating this as a low energy effective theory. In fact,
our theory does not have an explicit Planck scale in it so it is not even clear
what the cutoff energy is above which our theory is expected to break down.

Another issue is how to apply standard scalar-tensor theory limits 
from the Solar system, e.g.\ on the Jordan-Brans-Dicke parameter 
$\omega_{JBD}$.  Without the matter coupling and 
with $U=0$, one would say that $\omega_{JBD}=2\pi/\alpha$ 
and so these constraints \cite{bertotti} require 
$\alpha<10^{-4}$.  
In Ref. \cite{FiTrVe08} it is shown a compatibility  of the model $U=4$, $F=0$ 
with current cosmological parameters respecting Solar system constraints.   
However, 
the matter coupling might alter that conclusion.  
Ref.~\cite{PeBa08} considers the differences between matter coupled and 
nonminimally gravity coupled models, but so far no analysis has been 
carried out including both effects.  The equations 
for the field perturbations with all the couplings are quite complicated; 
we leave that analysis for future work and here only consider effects on 
the cosmic expansion.  

\begin{figure}[!htbp]
\includegraphics[width=\columnwidth]{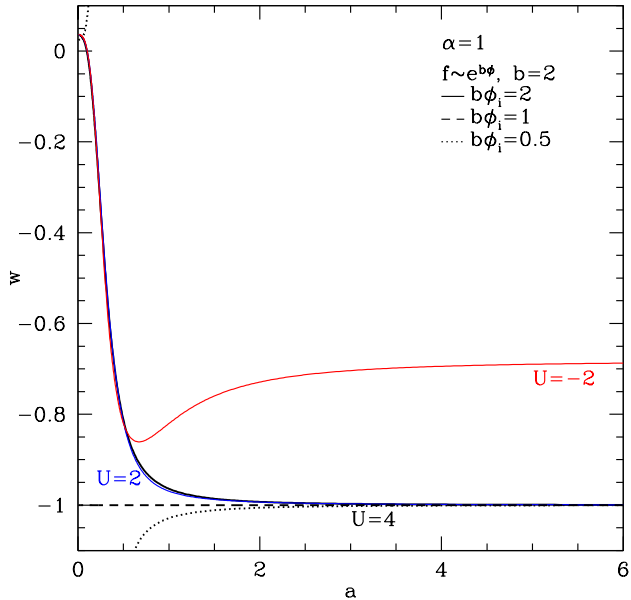}
\caption{Constant $U$ potentials with exponential coupling $f\sim e^{b\phi}$,  
shown for $b=2$.  For $U=4$ (thick, black curves), the cases for three 
initial field values 
$\phi_i$ are shown; for $U=2$ and $U=-2$ only the $b\phi_i=2$ case is 
plotted.  The $w(a)$ dynamics depends on $\phi_i$ (although it is quite 
similar to a corresponding constant $F$) but the final attractor 
depends on neither $f$ nor $\phi_i$.} 
\label{fig:varyf}
\end{figure}

We can consider going beyond constant $F$ and $U$.  
In the case of an exponential matter coupling, the logarithmic derivative 
$F$ is not constant.  Figure~\ref{fig:varyf} shows $w(a)$ for such a case, 
where $f\sim e^{b\phi}$, so $F=b\phi$ (here the field values are all in 
units of the Planck mass).  The 
dynamics is now dependent on the initial value of the field $\phi_i$, 
and resembles that of the constant $F$ case with $F=b\phi_i$ (a better 
approximation is evaluating this at a $\phi$ partway between $\phi_i$ 
and a later time value).  In particular this applies to the high 
redshift attractor.  However, note that the late time attractor given by 
Eq.~(\ref{eq:wc1}) remains the same, independent of $F$ and $\phi_i$.

\begin{figure}[!htbp]
\includegraphics[width=\columnwidth]{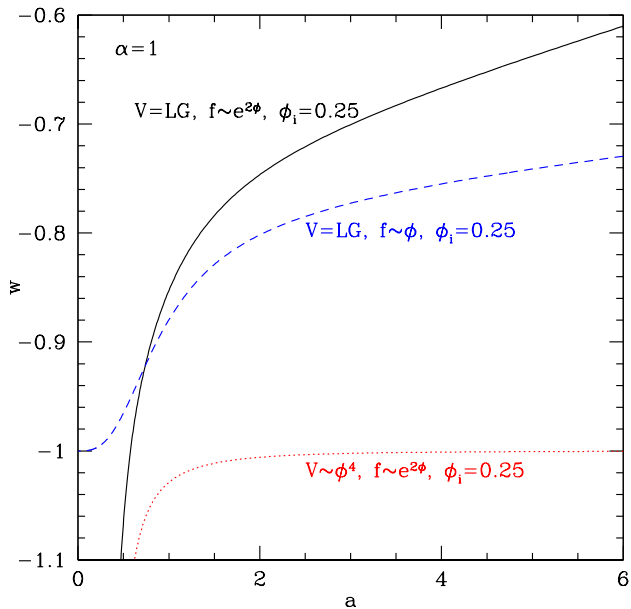}
\caption{Comparison of the Landau-Ginzburg (LG) form, with either constant 
$F$ or non power law $f$, to the $\phi^4$ potential shows significant 
differences.  In particular, the non power law potential does not exhibit 
attractor behavior.} 
\label{fig:varyv}
\end{figure}

One can also consider potentials where the logarithmic derivative $U$ 
is not constant.  
Figure~\ref{fig:varyv} illustrates $w(a)$ for a potential of 
Landau-Ginzburg form (\ref{langin}).  In this case, with varying $U$, 
the dynamics is quite different from the constant $U$ case, and no 
attractor solution is apparent.

\section{Conclusions} 

The dynamics of a scalar field model incorporating both nonminimal coupling 
to gravity and coupling to matter is rich.  The relations to induced 
gravity through a Higgs mechanism and the symmetrical ideas of coupling 
around the circle of scalar field, matter, and gravity are interesting. 
Furthermore, adding both couplings can relieve some problems of either 
individual coupling. 

A set of attractor solutions for the dynamics is found, with a stable 
attractor to a dark energy dominated universe with $w$ either equal to 
or near $-1$.  High redshift behavior is standard $\Lambda$CDM for a 
Yukawa matter coupling (better behaved than if there were no coupling), 
and a scaling solution otherwise.  Throughout cosmic history the matter 
and total equations of state can also be acceptably near $\Lambda$CDM 
behavior, for $\alpha$ not much larger than unity, depending on the couplings. 

A simple quartic potential $V\sim\phi^4$ appears viable (again better 
behaved than if the potential were constant), with a stable de 
Sitter attractor for all values of gravity and matter couplings.  Gravity 
is asymptotically restored to general relativity (at least as far as the 
background behavior is concerned).  A quadratic potential 
also has a de Sitter attractor. 

We identified the roles of the various couplings in the evolution 
equations and the effective potential.  
The slowing of field dynamics due to the gravity coupling likely 
alleviates problems with coupled matter instabilities, through the 
adiabatic mechanism discussed by \cite{Co08}, and the coupling to 
matter may help issues with gravity tests, but the system of perturbation 
equations becomes quite complicated due to the two extra scalar field 
couplings and we leave that to future work. 
Here we concentrated solely on the field and expansion dynamics, which 
in itself can constrain the parameter space, while pointing the way 
to interesting attractor behaviors and the possibility of dynamically 
establishing $w\approx-1$.

\acknowledgments

We gratefully acknowledge a UC-MEXUS-CONACYT Visiting Fellowship for JLCC 
to spend a sabbatical at Berkeley.  JLCC thanks the Berkeley Center for 
Cosmological Physics for hospitality and CONACYT for grant No. 84133-F.  
This work has been supported in part by the Director, Office of Science, 
Office of High Energy Physics, of the U.S.\ Department of Energy under 
Contract No.\ DE-AC02-05CH11231, and the World Class University grant 
R32-2009-000-10130-0 through the National Research Foundation, Ministry 
of Education, Science and Technology of Korea.



\begin{thebibliography}{499}
%
\bibitem{supernovae} 
A. G. Riess et al., Astron. J. {\bf 116}, 1009 (1998); \\ 
S. Perlmutter et al, ApJ, {\bf  517}, 565 (1999). 

\bibitem{union2} 
R. Amanullah et al., ApJ, {\bf 716}, 712 (2010).

\bibitem{wmap7} 
E. Komatsu et al, arXiv: 1001.4538. 

\bibitem{gal_surveys}  
B. A. Reid et al., MNRAS {\bf 404}, 60 (2010). 

\bibitem{uniqcmb} 
E.V. Linder and T.L. Smith, arXiv:1009.3500.

\bibitem{BrDi61} 
C. Brans and R. Dicke,  Phys. Rev.  {\bf 124}, 925 (1961). 

\bibitem{PeBaMa00}  F. Perrotta, C. Baccigalupi, and S. Matarrese, Phys. Rev. D {\bf 61}, 
023507 (2000). 

\bibitem{BaPi00}  N. Bartolo and M. Pietroni, Phys. Rev. D {\bf 61}, 023518 (2000).

\bibitem{BaMaPe00} 
C. Baccigalupi, S. Matarrese and F. Perrotta, Phys. Rev. D {\bf 62}, 123510 (2000).

\bibitem{Ch01} 
T. Chiba, Phys. Rev. D {\bf 64}, 103503 (2001).

\bibitem{FaJe06} V. Faraoni and M.N. Jensen, Classical Quantum Gravity {\bf 23}, 3005 (2006). 

\bibitem{Am00} L. Amendola,  Phys. Rev.  D {\bf 62}, 043511 (2000). 

\bibitem{DaCoKh06} 
S. Das, P.S. Corasaniti, and J. Khoury, Phys. Rev. D {\bf 73}, 083509 (2006). 

\bibitem{QuBrBaPi08}  
C. Quercellini, M. Bruni, A. Balbi, and D. Pietrobon, Phys. Rev. D {\bf 78},  063527 (2008). 

\bibitem{Co08} 
P.S. Corasaniti,   Phys. Rev.  D {\bf 78}, 083538 (2008). 
 
\bibitem{BeFlLaTr08} R. Bean, E.E. Flanagan, I. Laszlo, and M. Trodden, 
Phys. Rev. D {\bf 78}, 123514 (2008). 

\bibitem{ig_old} 
P. Minkowski, Phys. Lett. B {\bf 71},  419  (1977) ; 
A. Zee, Phys. Rev. Lett.  {\bf 42},   417  (1979);   
S.L. Adler, Rev. Mod. Phys. {\bf 54},  729 (1982) .                            

\bibitem{DeFrGh90s}  
H. Dehnen, H. Frommert, and F. Ghaboussi, Int. J. Theo. Phys. {\bf  29},   537 (1990); 
H. Dehnen and H. Frommert, Int. J. Theo. Phys. {\bf  30},  985 (1991). 

\bibitem{DeFrGh92} 
H. Dehnen, H. Frommert, and F. Ghaboussi, Int. J. Theo. Phys. {\bf  31}, 109 (1992). 

\bibitem{CeDe95ab} 
J. L. Cervantes-Cota  and H. Dehnen, Phys. Rev. D {\bf 51}, 395 (1995); \\
J. L. Cervantes-Cota  and H. Dehnen, Nucl. Phys. B {\bf 442}, 391 (1995). 

\bibitem{LiUr06} 
A. R. Liddle, L. A. Urena-Lopez,  Phys. Rev. Lett {\bf 97}, 161301 (2006). 

\bibitem{unifica06} 
T. Padmanabhan and T. R. Choudhury, 
Phys. Rev. \textbf{D66}, 081301(R) (2002);  R. J. Scherrer, 
Phys. Rev. Lett. \textbf{93}, 011301 (2004); A. Arbey, 
Phys. Rev. D{\bf 74}, 043516 (2006); P. J. E. Peebles and  A. Vilenkin, 
Phys. Rev. \textbf{D59}, 063505 (1999); J. E. Lidsey, T. Matos, and
L. A. Urena-Lopez, Phys. Rev. \textbf{D66}, 023514 (2002); T. Matos, 
J.-R. Luevano, and H. Garcia-Compean, \texttt{hep-th/0511098}. 

\bibitem{FaPe04}  G. R. Farrar and P. J. E. Peebles, ApJ,  {\bf 604}, 1 (2004).

\bibitem{DaPo94} T. Damour and A.M. Polyakov, Nucl. Phys. B {\bf 423}, 532 (1994); 

\bibitem{CoSaTs06} 
E. J. Copeland, M. Sami, S. Tsujikawa, Int. J. Mod. Phys. D {\bf 15} 1753 
(2006).

\bibitem{GRattractor} T. Damour and K. Nordtvedt, Phys. Rev. D {\bf 48}, 3436 (1993); 
D.I. Santiago, D. Kalligas and R.V. Wagoner, Phys. Rev. D {\bf 58}, 124005 (1998).

\bibitem{FiTrVe08} F. Finelli, A. Tronconi, and G. Venturi, Phys. Lett. B {\bf 659}, 466 (2008). 

\bibitem{Pe05} L. Perivolaropoulos,  JCAP {\bf 10}, 1 (2005). 

\bibitem{bertotti} 
B. Bertotti, L. Iess, and P. Tortora, Nature {\bf 425}, 374 (2003). 

\bibitem{PeBa08} V. Pettorino and C. Baccigalupi, Phys. Rev. D {\bf 77}, 
103003 (2008).

\end{thebibliography}
\end{document}